\documentclass[acmtog, balance = false, authorversion, nonacm]{acmart} 

\AtBeginDocument{%
  }

	\usepackage{wrapfig}
	\usepackage[table]{xcolor}
	\usepackage{mdframed}
	\usepackage{makecell}
	\usepackage{needspace}

\begin{document}

\title{Support-Free Planar Shell Printing of Arbitrary Shapes}
\author{Yefan Zhi}
\affiliation{%
	\institution{University of Pennsylvania}
	\city{Philadelphia}
	\state{PA}
	\country{USA}
}
\email{yefanzhi@design.upenn.edu}

\author{Masoud Akbarzadeh}
\affiliation{%
	\institution{University of Pennsylvania}
	\city{Philadelphia}
	\state{PA}
	\country{USA}
}
\email{masouda@design.upenn.edu}


\begin{abstract}
	Planar shell printing is a fabrication technique efficient in material usage and applicable to viscous materials.
	Prevalent in large-format 3D printing, it requires the toolpath to be support-free, where each shell layer must rest on the one below it.
	We present closed curve trimming, which directly optimizes the sliced toolpath curves of an arbitrary shape under a layer-wise containment constraint.
	By trimming each curve with the offset region of the layer below, our constructive algorithm guarantees the overhang limit and preserves the closed, consistent, and aligned (CCA) curve properties and the shell characteristics.
	The result is compatible with post-processing methods and continuous printing schemes.
	We demonstrate the algorithm on complex shapes, including TPMS and topology-optimized designs, with physical prints in polymer, clay, and concrete. It is implemented as a plugin for Rhino Grasshopper for accessible use.
\end{abstract}




\begin{teaserfigure}
	\includegraphics[width=\textwidth]{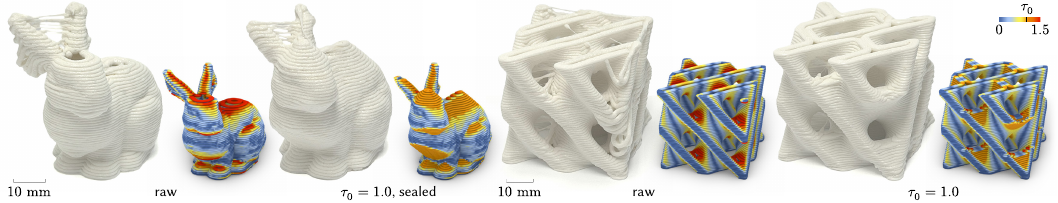}
	\caption{Closed curve trimming optimizes a toolpath so that it never exceeds the overhang limit. Coupled with overfill optimization and smoothing, it improves the buildability and surface quality of geometries with large overhangs. Examples printed using PLA with a $0.7\times 2.1 \textrm{ mm}$ extrusion profile.}
	\Description{figure description}
	\label{fig-teaser}
\end{teaserfigure}
\maketitle

\setlength{\intextsep}{4pt}
\setlength{\columnsep}{6pt}

\section{Introduction}

\definecolor{TypstGreen}{HTML}{2ecc40}
\definecolor{TypstRed}{HTML}{FF4136}
\definecolor{TabRowGray}{HTML}{EBEBEB}
\definecolor{TabOurs}{HTML}{FBCB8C}
\begin{table*}[t]
  \caption{We provide the first method that satisfies all optimization requirements.}
  \label{tab-current-methods}
  \centering
  \small
  \setlength{\tabcolsep}{4pt}
  \renewcommand{\arraystretch}{1.25}

  \newcommand{\Yes}{\textcolor{TypstGreen}{\ding{51}}}
  \newcommand{\No}{\textcolor{TypstRed}{\ding{55}}}
  \newcolumntype{L}[1]{>{\raggedright\arraybackslash\hyphenpenalty=10000\exhyphenpenalty=10000}m{#1}}
  \newsavebox{\CurrentMethodsBox}
  \sbox{\CurrentMethodsBox}{%
    \begin{tabular}{llcccccc}
      \toprule
      Optimization method         & Literature                                                                                  & \makecell[c]{Arbitrary \\ shape} & \makecell[c]{Preserved \\ shape} & \makecell[c]{Planar shell \\ printing} & \makecell[c]{Guaranteed \\ bound} & Venue   & \makecell[c]{Material \\ strategy} \\
      \midrule
      Voxel topology optimization & \cite{Gaynor-2016-3DP-Topological-Optimization,Langelaar-2016-3DP-Topological-Optimization} & \No                              & \No                              & \No                                    & \Yes                              & volume  & generate                           \\
      \rowcolor{TabRowGray}
      Morphable component         & \cite{Bhat-2022-3DP-Topological-Optimization,Guo-2017-3DP-Topological-Optimization}         & \No                              & \No                              & \Yes                                   & \Yes                              & volume  & modify                             \\
      Multi-axis printing         & \cite{Zhang-2022-Multi-Axis-Printing,Dai-2018-Multi-Axis-Printing}                          & \Yes                             & \Yes                             & \No                                    & \Yes                              & volume  & modify                             \\
      \rowcolor{TabRowGray}
      Shape optimization          & \cite{Hu-2015-Overhang,Zhi-2025-Overhang-2}                                                 & \Yes                             & \Yes                             & \Yes                                   & \No                               & surface & modify                             \\
      Curve generation            & \cite{Zhong-2020-Overhang,Mogra-2023-Overhang}                                              & \No                              & \No                              & \Yes                                   & \Yes                              & curve   & generate                           \\
      \rowcolor{TabRowGray}
      Curve addition              & \cite{Tricard-2020-Support-Structure,Xing-2021-Overhang}                                    & \Yes                             & \No                              & \Yes                                   & \Yes                              & curve   & add                                \\
      Voxel sculpting             & \cite{Reiner-2016-Overhang}                                                                 & \Yes                             & \Yes                             & \No                                    & \Yes                              & layer   & subtract/add                       \\
      \rowcolor{TabOurs}
      Closed curve trimming       & Ours                                                                                        & \Yes                             & \Yes                             & \Yes                                   & \Yes                              & curve   & subtract/add                       \\
      \bottomrule
    \end{tabular}}
  \usebox{\CurrentMethodsBox}

  \vspace{0.6ex}
  \parbox{\wd\CurrentMethodsBox}{\footnotesize\raggedright
    Preserved shape: after optimizing an arbitrary input, the overall shape and topology are kept, without adding artificial features such as ribs or skeletons.}

\end{table*}

In conventional layer-based material extrusion 3D printing, geometric features such as large overhangs, long bridges, and big holes can only be produced successfully with tailored support structures \cite{Vanek-2014-Support-Structure,Dumas-2014-Support-Structure,Schmidt-2014-Support-Structure}. The support structures are typically printed together with the main body and removed afterward.

This paper focuses on planar shell printing, which is prevalent in large-format centimeter- to meter-scale 3D printing practices due to its efficient material use and compatibility with viscous materials including concrete, clay, and glass.
In these scenarios, support structures significantly increase material and energy waste. The viscous nature of the material also makes the supports difficult to remove, especially when internal support is required for functional components with porous shapes.
These downsides of support structures have led to the \emph{support-free} requirement for large-format planar shell printing.

\paragraph{Why planar shell printing}
In large-scale construction practices, 5- or 6-axis robotic arms are not widely available due to their cost and restrained scale, limiting the feasibility of multi-axis printing \cite{Xiao-2021-3DP-Concrete-Review}.
On the other hand, concrete and other paste materials involve a slow hydration process and carry non-negligible self-weight \cite{Breseghello-2022-Adaptive-Extrusion}. To prevent slumping and breaking, paste materials are mostly printed in planar or slightly curved layers which remain feasible on 3-axis gantries \cite{Etienne-2019-Non-Parallel-3DP}.
For efficient material use, expressive design, and application as formwork, 3D printing for construction purposes typically prints only the hollow shell surface of a structurally/functionally relevant geometry, rather than the entire solid volume \cite{Bos-2016-3DP-Concrete-Review}.


\paragraph{Support-free planar shell printing}
This paper presents the \emph{closed curve trimming} method that optimizes a toolpath for support-free planar shell printing.
It applies to arbitrary manifold shapes with closed toolpath curves, operates on the toolpath level, and reduces extreme overhang. The result preserves the properties of planar shell printing, guarantees the overhang limit, and is compatible with continuous printing planning algorithms.
Intrinsically, the algorithm is designed to optimize ``mostly buildable'' geometries. However, we show that in certain cases, when applied to extremely unbuildable shapes, it simplifies intricate features and makes the toolpath support-free while preserving the overall shape and topology of the design.

\Needspace{3\baselineskip}
Our contributions are:
\begin{itemize}
	\item A curve-level formulation of overhang optimization for planar shell printing, in which the input and output are layers of closed, consistent, and aligned (CCA) curves, and buildability is enforced as a layer-wise containment constraint.
	\item A constructive closed curve trimming algorithm built on curve offsetting and polygon clipping, which guarantees the overhang bound while preserving the shell nature and the CCA properties.
	\item Physical validation across scales with polymer, clay, and concrete materials.
	\item Case studies on triply periodic minimal surfaces (TPMS) and topology-optimized designs  for extended engineering applications.
	\item Implementation as a plugin for Rhino Grasshopper for accessibility across disciplines.
\end{itemize}

Although motivated by large-format 3D printing, the planar shell printing problem we address is purely geometric. The method is independent of material and scale, and applies equally to desktop fused filament fabrication.
It complements existing support-free fabrication techniques in the graphics community and can be passed directly to continuous toolpath planning methods.
More broadly, the work shows how classical polygon offsetting and clipping methods, when made robust to the singularities, yield a simple and practical tool for design-for-fabrication.

\section{Related Work}
Before we offer formal definitions of overhang and its optimization problem for planar shell printing, we review related work on support-free fabrication.

\paragraph{Generative methods}
Voxel-based topology optimization uses voxel grids or density fields to generate efficient structures under given boundary conditions
and can be combined with overhang constraints \cite{Gaynor-2016-3DP-Topological-Optimization,Langelaar-2016-3DP-Topological-Optimization}.
Morphable-component-based topology optimization represents the overall optimized shape as an aggregation of elements with explicit geometries where overhang angle can be directly regulated \cite{Bhat-2022-3DP-Topological-Optimization, Guo-2017-3DP-Topological-Optimization}.
While these generative methods are effective in generating support-free, structurally efficient designs, they cannot apply to arbitrary shapes and are thus limited in applicability.

\paragraph{Shape-level methods}
A widely studied method in polymer 3D printing that creates support-free printing schemes is multi-axis printing, where the design is printed in curved layers with varying nozzle orientation \cite{Dai-2018-Multi-Axis-Printing,Li-2022-Path-Orientation,Zhang-2022-Multi-Axis-Printing,Fang-2020-Multi-Axis-Printing,Qu-2026-Multi-Axis-Printing}.
By creating larger contact area between layers to enhance friction and relying on the intra-layer bonding to counter gravity, extremely unbuildable designs can be printed without support structures.
However, this method is largely exclusive to polymer materials which solidify instantly upon extrusion.

Another trend optimizes the global orientation of the object to reduce overhangs and the need for support structures \cite{Zhang-2015-Support-Structure,Jayakody-2025-Multi-Axis-Printing}. Such schemes easily create concentration of stress. For paste materials, this can lead to significant sagging and deformation due to their self-weight and viscoelastic behavior.

A more practical approach in large-format printing is to optimize the shape such that overhangs are reduced in planar printing with respect to the given printing direction. This can be done through regulating the surface normals \cite{Hu-2015-Overhang,Zhi-2025-Overhang-2,Liu-2021-Shape-Stylization} or optimizing layers as a wireframe model \cite{Zhi-2023-Overhang}. However, these iterative methods treat overhang as a soft penalty and thus cannot guarantee the bound; moreover, surface normals alone are insufficient to ensure buildability (example in Figure~\ref{fig-examples-topoopt}).

\paragraph{Curve/layer-level methods}
Some works in large-format shell printing directly generate support-free toolpath curves based on sections \cite{Zhong-2020-Overhang} or layer evolution \cite{Mogra-2023-Overhang}, achieving support-free printing in restricted design spaces.

Additive post-processing methods apply to given toolpaths and add permanent support structures such as ribs \cite{Tricard-2020-Support-Structure} or additional loops \cite{Xing-2021-Overhang}.
Support-free hollowing reduces the material use in solid models \cite{Wang-2018-Support-Structure,Mao-2018-Support-Structure,Zhang-2015-Support-Structure-2}.
However, for large-format printing, these methods add substantial material to the design and change its shell nature.

One promising method that preserves the shape characteristics is voxel sculpting \cite{Reiner-2016-Overhang}, where the geometry is seen as layers of voxels. Unsupported voxels are trimmed and required supports are added. The operations propagate through the layers to guarantee buildability. However, the voxel-based approach is not applicable to shell printing.

\paragraph{Ours}
We introduce closed curve trimming, a constructive method that directly modifies the toolpath curves by trimming them with the offset region of the previous layer, in a single bottom-up pass over the layers.
As shown in Table~\ref{tab-current-methods}, it is the first method that satisfies all the optimization requirements: It applies to arbitrary shapes, preserves the shell shape characteristics, supports planar shell printing, and guarantees the overhang bound.

We adopt the overhang definition of \citet{Zhi-2025-Continuous-Path} and the support-free definition of \citet{Hornus-2015-Overhang}.
The method extends the subtraction method of \citet{Reiner-2016-Overhang} to apply to shell printing.
The subtraction is done through closed curve trimming followed by post-processing steps to preserve surface integrity and curve continuity, where new curve segments are added deterministically if necessary, making the result a combination of subtraction and addition.

\section{Overhang optimization problem}\label{sec-overhang-optimization-problem}

The proposed algorithm optimizes a sliced set of toolpath curves. It is thus crucial to define a compatible toolpath, the overhang property, and the objective.

\subsection{Closed, consistent, and aligned (CCA) toolpath curves}\label{sec-closed-toolpath-curves}

We adopt the toolpath data structure of \citet{Zhi-2025-Continuous-Path}, where any arbitrary 3D shape that is manifold and closed, when sliced, is represented as a sequence of planar layers with closed, non-intersecting curves (polygons). The adjacent layers are parallel or have a small angle. We summarize three properties of the toolpath curves that are naturally established in such structures and should be preserved in the proposed method:

\begin{figure}
	\centering
	\includegraphics[width=85mm]{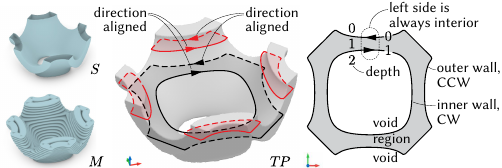}
	\caption{Well-oriented, closed and non-intersecting curves form regions with aligned orientation across layers. Example of a double-sided Schwarz P triply periodic minimal surface (TPMS).}
	\Description{}
	\label{fig-orientation-and-region}
\end{figure}

\paragraph{Closed and non-intersecting}
Closed and non-intersecting curves represent the manifold boundaries of a 2D solid volume, which is the result of slicing a 3D solid volume.
Appendix~\ref{apd-touching-boundary} discusses the singularity of point intersections in representing the volume.

\paragraph{Consistent orientation}
Intrinsically, a parametric curve has a direction, which also directs the movement of the extruder in 3D printing. Closed and non-intersecting curves can be consistently \emph{oriented}. We follow the convention to orient the curves counterclockwise (CCW) to create positive volumes and clockwise (CW) to create negative volumes \cite{ORourke-1998-Computational-Geometry,De-Berg-2008-Computational-Geometry,OGC-2011-Polygon}. For a polygon volume with holes, the \emph{outer wall} is CCW and the \emph{inner walls} are CW (when seen from above, from the $+ z$ direction). We refer to the continuous 2D volume enclosed by these wall curves as a \emph{region}. The consistent orientation scheme gives an important property to the curves---the left side of a curve is always the interior of the region. This link between the local characteristic of curve direction and the global property of curve closedness is crucial for polygon clipping, offsetting, and our method of closed curve trimming.

Figure~\ref{fig-orientation-and-region} shows an example of such orientation scheme. The wall curves form a hierarchical structure of enclosure, where some curves enclose other curves, which in turn may enclose more curves, etc.\ \citep[see][]{OpenCV-2016-Polygon}. Vertically along the hierarchy, the orientation of the curves alternates between CCW and CW. We can thus use \emph{depth} to determine the orientation of the curves and the associated regions, which is defined as the number of (other) curves that enclose a curve/point. A curve is a CCW outer wall if its depth is even. A point is in a region if its depth is odd. With consistent orientation, the region will have a \emph{winding number} of $1$ and the voids will have a winding number of $0$.

\paragraph{Aligned orientation across layers}
Figure~\ref{fig-orientation-and-region} also shows that for a sliced toolpath, the consistent orientation scheme aligns curve directions across layers, even though the type of wall (inner/outer) might not match. The alignment is compatible with continuous printing planning algorithms that create ramps across layers to minimize the number of stop-and-starts in extrusion \citep[e.g.,][]{Zhi-2025-Continuous-Path,Zhong-2022-Continuous-Path,Bi-2022-Continuous-Path}. It also gives a crucial criterion for connecting trimmed segments in the post-processing step of our method. Aligned orientation across layers is the natural result of slicing a solid volume. However, for a toolpath without a reference volume, there is not a clear and operational definition for such matching and alignment. For a toolpath with little correlation between adjacent layers, the alignment cannot be assessed. Therefore, in this paper, we assume alignment is given by the consistent orientation scheme and is preserved if, after a sliced toolpath is modified, the segments preserved will still have their orientation unchanged.

In short we say a toolpath is ``CCA'' if its curves are closed and non-intersecting, consistently oriented, and aligned across layers.

\subsection{Local overhang}

Overhang takes place when material reaches out horizontally without being adequately supported. To assess the buildability of a printing scheme, \citet{Zhi-2025-Continuous-Path} have proposed measuring two types of overhang: \emph{local overhang} between adjacent layers to avoid plastic collapse and \emph{global overhang} of the mass across multiple layers to prevent elastic buckling. These criteria are purely geometric and independent of material properties and printing parameters such as speed and time, which makes them handy tools for design communication. For a specific material recipe and the associated printing conditions (temperature and humidity), standard tests could be performed to record printable upper bounds of the two types of overhangs for reference.

This paper focuses on the local overhang, since it can be effectively optimized with small local modifications, while the global overhang is more related to the overall shape and topology of the design and usually requires more significant adjustments to be optimized. Hereafter we will refer to local overhang as simply overhang.

\begin{figure}
	\centering
	\includegraphics[width=85mm]{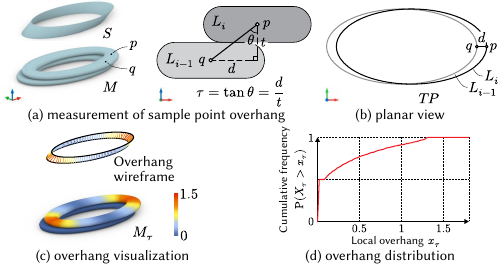}
	\caption{
		Overhang can be measured using sample points and visualized as a color-coded mesh. Cumulative frequency allows the quantitative comparison between overhang distributions.
	}
	\Description{}
	\label{fig-definition-of-overhang}
\end{figure}

Local overhang assesses the support a toolpath point receives from the previous layer.
Figure~\ref{fig-definition-of-overhang} illustrates two adjacent toolpath layers $L_{i - 1}$ and $L_{i}$ with a layer thickness of $t$ (For slightly angled layers, $t$ is the nominal or average thickness). For any toolpath point $p \in L_{i}$, find $q \in L_{i - 1}$ from the previous layer that is closest to $p$'s projection on $L_{i - 1}$. $q$ is seen as the support point of $p$. The \emph{overhang distance} $d$ of $p$ is the distance between $q$ and $p$'s projection on $L_{i - 1}$. It is the planar offset of $p$ from its support point. The dimensionless \emph{overhang} $\tau$ is the overhang distance divided by the layer thickness, $\tau = d/t$, or the tangent of the overhang angle $\tan\theta$. In the rest of the paper, we adopt a planar view as in Figure~\ref{fig-definition-of-overhang}b, where all layers are placed in the same ${\mathbb{R}}^{2}$ space by matching their base planes\footnote{In parallel/non-parallel planar layers, base planes between adjacent layers are aligned such that the next plane is the previous plane translated along its $+ z$ direction or rotated about an axis on the plane. This rule prevents unnecessary distortion. Layer thickness $t$ is measured as the length of the trajectory of the plane.

	For non-planar layers, if there exists a bidirectional mapping strategy that converts them to planar layers while preserving the relative positions of points, our method can be applied to enforce an approximate overhang constraint.
}. The overhang distance of $p \in L_{i}$ is
\begin{align}
	d = \min\left\{ \left\| {p,q} \right\|~|~ q \in L_{i - 1} \right\}.
\end{align}
By generating equally spaced sample points on the toolpath curves and connecting them with their support points, we can visualize the overhang on a toolpath as a wireframe model. By color-coding the printing visualization mesh $M$, we can also visualize the overhang as $M_{\tau}$. The distribution of overhang $X_{\tau}$ can be represented by the cumulative frequency $y = \mathrm{P}(X_{\tau} > x_{\tau})$ for comparison.

\begin{figure}
	\centering
	\includegraphics[width=85mm]{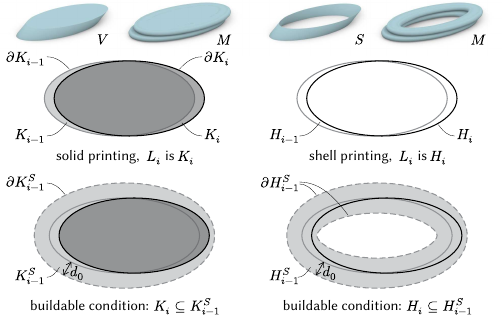}
	\caption{A buildable layer in solid/shell printing requires $L_{i} \subseteq L_{i - 1}^{S}$.}
	\Description{}
	\label{fig-a-buildable-layer}
\end{figure}

\subsection{Objective}

Given an overhang distance limit $d_{0} = \tau_0 t$ (inclusive), all possible points that can be supported by $L_{i - 1}$ form a set
\begin{align}
	L_{i - 1}^{S} = \left\{ p \in {\mathbb{R}}^{2}~|~\exists q \in L_{i - 1},\left\| {p,q} \right\| \leq d_{0} \right\}.
	\label{eq-offset-both-plain}
\end{align}
$L_{i}$ is fully supported if the containment condition holds:
\begin{align}
	L_{i} \subseteq L_{i - 1}^{S}.
\end{align}
Let $C_{d_{0}}$ denote a zero-centered disk area with $d_{0}$ radius (inclusive).
Eq.~(\ref{eq-offset-both-plain}) is equivalent to the Minkowski sum
\begin{align}
	L_{i - 1}^{S} = L_{i - 1} \oplus C_{d_{0}} = \left\{ q + r~|~ q \in L_{i - 1}, r \in C_{d_{0}} \right\}.
	\label{eq-offset-minkowski-plain}
\end{align}
It is also the region formed by offsetting (dilating) the boundary curves of $L_{i - 1}$ by distance $d_{0}$ with round corners. This is consistent with the support-free definition by \citet{Hornus-2015-Overhang}.

Recall the CCA requirements, we can formalize the overhang optimization problem as the following:

\begin{mdframed}[
		linewidth=0.8pt,
		linecolor=black,
		backgroundcolor=white,
		innerleftmargin=4pt,
		innerrightmargin=4pt,
		innertopmargin=4pt,
		innerbottommargin=4pt,
		skipabove=6pt,
		skipbelow=6pt
	]
	\textbf{Overhang Optimization Problem.}
	For a CCA toolpath with $n$ layers $L_{0..n - 1}$, find modified layers $L'_{0..n - 1}$ such that, for ${i = 1,\ldots,n - 1}$,
	\begin{align}
		L'_{i} & \subseteq L'^{S}_{i - 1},         \label{eq-overhang-optimization-problem}
	\end{align}
	and the new curves are still CCA.
\end{mdframed}

\vspace{1ex}

\paragraph{Preserving the shell shape} The above formulation is actually a weak feasibility problem without an objective function for optimization.
In practice, we aim to preserve the shape and its shell characteristics with minimal modifications. However, it is open-ended to define and measure the preservation of the shell characteristics and the minimality of modifications.
Our method satisfies Eq.~(\ref{eq-overhang-optimization-problem}) and the CCA properties by construction, while shape preservation is pursued heuristically through two empirical constraints that yield conservative shape modifications.
For the two adjacent layers $L_{i - 1}$ and $L_{i}$, assume the optimized $L'_{i - 1}$ is obtained and preserves the shell characteristics\footnote{Eqs.~(\ref{eq-local-modification-1}) and (\ref{eq-local-modification-2}) might not be achievable if the supported region is evaluated on the original $L_{i - 1}$ since it might have to be changed drastically.}.
Firstly, if a part in layer $L_i$ is supported by the optimized version of its previous layer $L'_{i - 1}$, it should remain unchanged:
\begin{align}
	L_i \cap L'^{S}_{i - 1} \subseteq L'_i.
	\label{eq-local-modification-1}
\end{align}
Secondly, added parts, if any, should be contained in the boundary of the supported region $\partial L'^{S}_{i - 1}$, whose shape is determined by offsetting the bottom layer $L'_{i - 1}$ and thus deemed consistent with the two layers:
\begin{align}
	L'_i \setminus L_i \subseteq \partial L'^{S}_{i - 1}.
	\label{eq-local-modification-2}
\end{align}
Naturally, artificial features like ribs or skeletons will be avoided. However, since we ensure these conditions sequentially across layers, deviations from the original layers may accumulate and the result can smooth out some fine details of the original shape (see Section~\ref{sec-blending-and-smoothing-effects}).


\section{Closed curve trimming method}\label{sec-closed-trimming-method}

The optimization method of shell printing is an extension of the primitive condition of solid printing. Therefore we introduce the method in both scenarios. To differentiate, we write $L_{i}$ as $K_{i}$ in solid printing and $H_{i}$ in shell printing (Figure~\ref{fig-a-buildable-layer}). In both conditions we start with $L'_{0} \leftarrow L_{0}$ and optimize the layers sequentially from bottom to top: Each time we assess two adjacent layers $L'_{i - 1}$ and $L_{i}$; The bottom layer is already optimized to be buildable and thus fixed, while the upper layer $L_{i}$ is modified. The rest of this section therefore focuses on two adjacent layers. For simplicity we write the fixed bottom layer as $L_{i - 1}$.

\subsection{Solid printing}

\begin{figure}
	\centering
	\includegraphics[width=85mm]{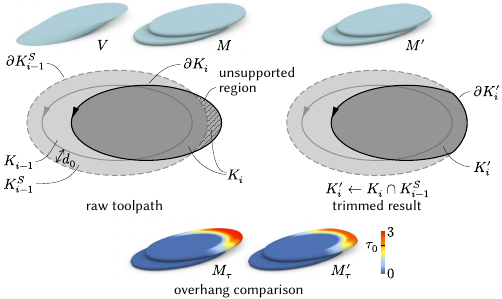}
	\caption{Closed curve trimming for overhang optimization in solid printing is essentially 2D region intersections.}
	\Description{}
	\label{fig-solid-printing}
\end{figure}

In solid printing of a volume $V$, we express a toolpath layer as a solid region $K$.
Its boundary is a collection of closed, non-intersecting curves $\partial K = \left( c_{0},c_{1},\ldots \right)$, which are the result of slicing the surface $S$ of the solid volume $V$ and naturally CCA. In the solid printing scenario, we assume the region can be properly filled and only enforce the CCA properties of the boundaries.

As illustrated in Figure~\ref{fig-solid-printing}, following Eq.~(\ref{eq-local-modification-1}), the natural strategy to ensure $K_{i} \subseteq K_{i - 1}^{S}$ is to trim out the unsupported region and keep the intersection of $K_{i}$ and $K_{i - 1}^{S}$,
\begin{align}
	K'_{i} \leftarrow K_{i} \cap K_{i - 1}^{S}.
	\label{eq-trimming-solid-plain}
\end{align}
This is seen in \citet{Reiner-2016-Overhang} where layers are represented as voxels\footnote{Their method also includes a top-down grow operation that adds material to create ``closest supporting paths.'' We exclude it here since it does not extend to shell printing.}. As we represent layers by their boundary curves, it becomes a polygon clipping operation that can be efficiently computed with graph-based algorithms \cite{Weiler-1977-Polygon,Vatti-1992-Polygon,Gong-2009-Polygon} and libraries \cite{clipper2,Fogel-2026-Polygon}. In the example of Figure~\ref{fig-solid-printing}, the boundary $\partial K_{i}$ is trimmed by the boundary of the supported region $\partial K_{i - 1}^{S}$. Since the result is still a collection of solid regions and the orientations of the unchanged segments are preserved (the left is still the interior), CCA properties are preserved.
Since no material is added, Eq.~(\ref{eq-local-modification-2}) holds.
Let $V$ and $V'$ represent the solid 3D volume printed before and after trimming all layers. This strategy always has
\begin{align}
	V' \subseteq V
\end{align}
and thus we say it uses a subtractive material strategy.

\subsection{Shell printing}

In shell printing of a surface $S$, we express a toolpath layer as a collection of closed, non-intersecting curves $H = \left( c_{0},c_{1},\ldots \right)$. We use a minimal example to illustrate the key idea of the optimization in Section~\ref{sec-minimal-example} and provide a complete and operational algorithm in Section~\ref{sec-complete-algorithm}.

\subsubsection{Minimal example}\label{sec-minimal-example}

\begin{figure}
	\centering
	\includegraphics[width=85mm]{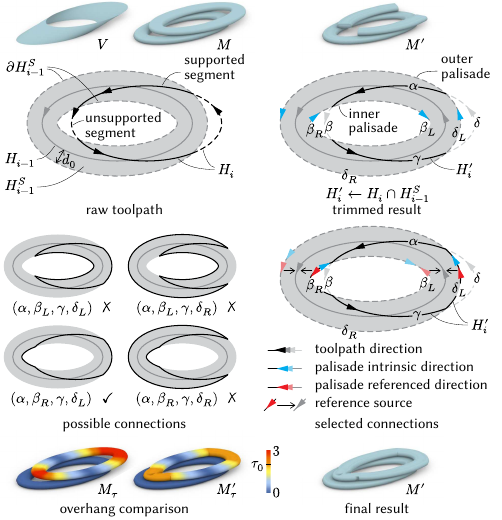}
	\caption{Closed curve trimming for overhang optimization in shell printing selects palisade connections based on their referenced directions.}
	\Description{}
	\label{fig-shell-printing}
\end{figure}

In the example of Figure~\ref{fig-a-buildable-layer}, $H_{i - 1}^{S}$ is a ring region with one outer wall and one inner wall. We refer to the closed curves of its boundary $\partial H_{i - 1}^{S}$ as \emph{palisades}. Consider adopting the trimming strategy in Eq.~(\ref{eq-trimming-solid-plain}), $H'_{i} \leftarrow H_{i} \cap H_{i - 1}^{S}$. For a layer that is not fully supported, $H'_{i}$ will include open curves. In the example of Figure~\ref{fig-shell-printing}, the curve in the upper layer $H_{i}$ is split into four segments $(\alpha,\beta,\gamma,\delta)$ by the palisades. Only $\alpha$ and $\gamma$ are supported and preserved. This breaks the continuity and closedness required. The natural strategy is to connect $\alpha$ and $\gamma$ with curves inside the region of $H_{i - 1}^{S}$.
Eq.~(\ref{eq-local-modification-2}) suggests that we use palisade segments in $\partial H_{i - 1}^{S}$ for this purpose. The illustration shows that, when properly selected, they suggest minimal deviation from the trimmed input.

Take $\beta$ as an example: it splits the inner palisade into two segments, $\beta_{L}$ on its left side, and $\beta_{R}$ on its right side. When we try to extend the end of $\alpha$, we can choose $\beta_{L}$ or $\beta_{R}$. They both lead to the start of $\gamma$ and are thereby candidates to substitute $\beta$. Similarly, $\gamma_{L}$ and $\gamma_{R}$ are candidates to substitute $\gamma$. Judging by eye, of all four possibilities to form a complete closed curve, we know that $\left( \alpha,\beta_{R},\gamma,\delta_{L} \right)$ is the best choice, as it is closest to both the original curve and a reference well-supported condition illustrated in Figure~\ref{fig-a-buildable-layer}.

To offer a generalized criterion for selecting the best candidate, we look at the orientation of the palisade curves.

\begin{itemize}
	\item
	      Since they form regions, they have \emph{intrinsic directions} (blue) following which the left side is always the interior. They can be solved as in Section~\ref{sec-closed-toolpath-curves} and are the default result of certain computational geometry libraries.
	\item
	      Since they are offset results from the skeleton of $H_{i - 1}$, they also have \emph{referenced directions} (red) parallel to the skeleton. The thin arrows in Figure~\ref{fig-shell-printing} show the reference source of the palisade segments.
\end{itemize}

The two directions can be either the same (as for $\delta_{L}$) or opposite (as for $\beta_{R}$).
The example of Figure~\ref{fig-shell-printing} suggests that if we solve the referenced direction of all segments and select those consistent with the supported toolpath segments ($\alpha$ and $\gamma$) after connection, we can get the desired solution.

\subsubsection{Complete algorithm}\label{sec-complete-algorithm}

\begin{figure}
	\centering
	\includegraphics[width=84.875mm]{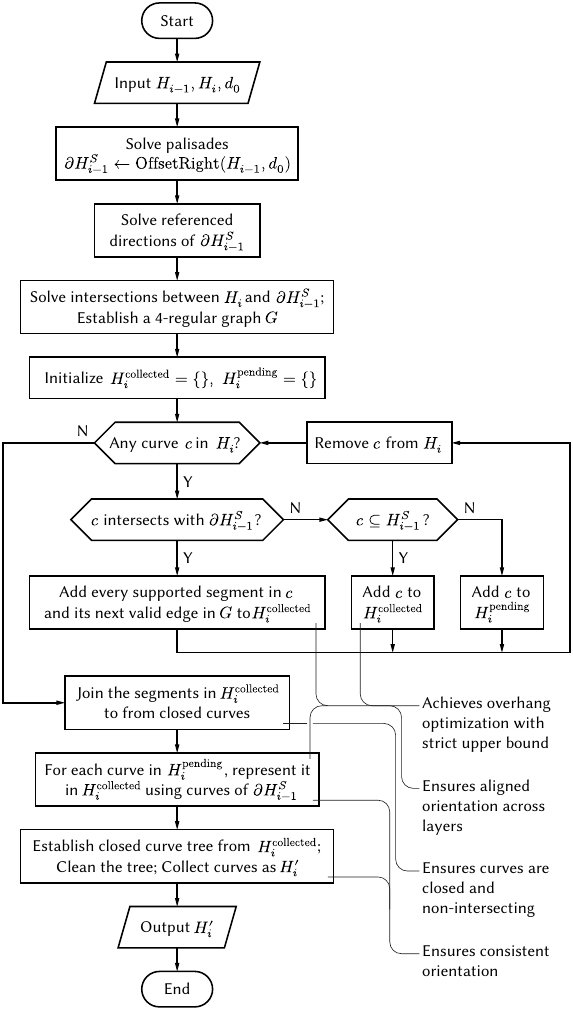}
	\caption{The complete algorithm for overhang optimization in shell printing achieves the objective and preserves the CCA properties.}
	\Description{}
	\label{fig-flowchart}
\end{figure}

The complete algorithm, illustrated in Figure~\ref{fig-flowchart}, offers operational criteria for all corner cases. We break down the steps here and offer additional details in the following subsections.

\begin{enumerate}
	\item
	      Take as input the two adjacent layers $H_{i - 1}$ and $H_{i}$ as well as the overhang distance limit $d_{0}$.
	\item
	      Solve the palisades, which are the results of offsetting the directed curves in $H_{i - 1}$ by distance $d_{0}$ to their right side:
	      \begin{align}
		      \partial H_{i - 1}^{S} \leftarrow \text{ OffsetRight}\left( H_{i - 1},d_{0} \right).
	      \end{align}
	      Find their intrinsic directions and referenced directions (Section~\ref{sec-referenced-direction}).
	\item
	      Solve intersections between $H_{i}$ and the palisades. Establish a 4-regular graph $G$ with the edges from the two sources (Section~\ref{sec-supported-segments}).
	\item
	      Maintain two sets, $H_{i}^{\text{collected}}$ for collected edges and curves, and $H_{i}^{\text{pending}}$ for pending ones. For each curve $c$ in $H_{i}$, there are three possibilities:

	      \begin{itemize}
		      \item
		            $c$ intersects with the palisades.
		            $c$ will be split into at least 2 segments in $G$. (Seen as not intersecting if there is only one point intersection.) For each segment, if it is supported, add it and its next valid edge in $G$ to $H_{i}^{\text{collected}}$ (Section~\ref{sec-supported-segments}).
		      \item
		            $c$ does not intersect with the palisades. Solve its depth in $H_{i - 1}^{S}$. It is fully supported if it has an odd depth.

		            \begin{itemize}
			            \item
			                  If it is fully supported, $c \subseteq H_{i - 1}^{S}$. Add it to $H_{i}^{\text{collected}}$.
			            \item
			                  If it is not fully supported, add it to $H_{i}^{\text{pending}}$ for further processing to ensure consistent orientation.
		            \end{itemize}
	      \end{itemize}
	\item
	      Join the segments in $H_{i}^{\text{collected}}$. The results will be non-intersecting closed curves only (proven in Section~\ref{sec-supported-segments}).
	\item
	      For each curve $c$ in $H_{i}^{\text{pending}}$, find its parallel adjacent curve(s) in $\partial H_{i - 1}^{S}$. If they are not represented in $H_{i}^{\text{collected}}$, add them to $H_{i}^{\text{collected}}$ to ensure consistent orientation (Section~\ref{sec-replace-unsupported-curves}).
	\item
	      Establish a closed curve tree from $H_{i}^{\text{collected}}$. Clean it by removing curves breaking the consistent orientation. Collect the curves in the cleaned tree as the final result $H'_{i}$ (Section~\ref{sec-clean-final-result}).
	\item
	      Output the modified layer $H'_{i}$.
\end{enumerate}

\subsubsection{Referenced direction}\label{sec-referenced-direction}

\begin{figure}
	\centering
	\includegraphics[width=85mm]{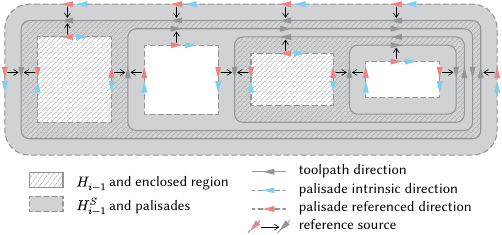}
	\caption{The referenced direction of a palisade curve is flipped from its intrinsic direction if it has an odd depth in $H_{i - 1}$ and is thus inside the region.}
	\Description{}
	\label{fig-direction}
\end{figure}

Referenced directions can be solved in the following way:

\begin{itemize}
	\item
	      If a segment is on the right side of the directed skeleton (toolpath curve from the previous layer), its left should be the interior of the region $H_{i - 1}^{S}$ which contains the skeleton. Therefore, its referenced direction is the same as its intrinsic direction.
	\item
	      If a segment is on the left side of the directed skeleton, its right should be the interior of the region $H_{i - 1}^{S}$ which contains the skeleton. Therefore, its referenced direction is the opposite of its intrinsic direction.
\end{itemize}

A segment is on the left side of the directed skeleton if and only if it is inside the region formed by the skeleton curves $H_{i - 1}$. Since $\partial H_{i - 1}^{S}$ is the boundary of the region formed by offsetting the curves in $H_{i - 1}$, the two sets of closed curves will not intersect, and the referenced directions for palisade segments are consistent in each palisade curve, which can be determined by evaluating that curve's depth in $H_{i - 1}$. In one sentence, the referenced direction of a palisade curve is flipped from its intrinsic direction if and only if it has an odd depth in $H_{i - 1}$. Figure~\ref{fig-direction} gives a complex example that validates this simple criterion.

\subsubsection{Curve trimming with palisade connections}\label{sec-supported-segments}

This subsection discusses the operational steps for trimming the toolpath curves partly supported and connecting them with palisade connections using a graph-based approach, briefly shown in the minimal example in Section~\ref{sec-minimal-example}. It only concerns those toolpath curves that intersect with the palisades.

\Needspace{12\baselineskip}
By solving the intersections between the toolpath curves and the palisades, we form a 4-regular graph $G$ with edges from the two
\begin{wrapfigure}{r}{30mm}
	\centering
	\includegraphics[width=30mm]{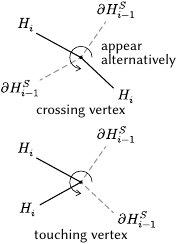}
\end{wrapfigure}
sources. Say the edges are therefore of two types. Every vertex in $G$ is a crossing vertex, where the two types of edges appear alternately in CCW order. (Touching vertices are ignored in $G$. Overlapping singularities are handled in Appendix~\ref{apd-overlap}.) Such a graph can be processed vertex-by-vertex to compute polygon clipping operations, as shown in \cite{Weiler-1980-Polygon,Gong-2009-Polygon}. Here we use the vertex-level information to select the supported segments and proper palisade connections.

\begin{figure}
	\centering
	\includegraphics[width=85mm]{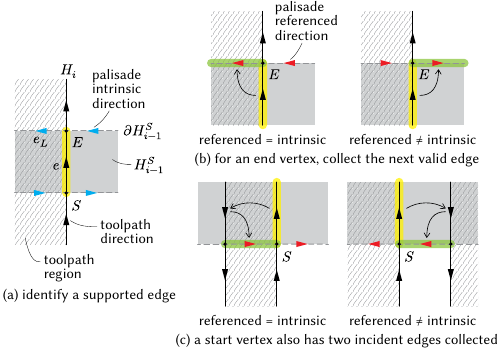}
	\caption{From vertex-level information, supported edges and their next valid edge can be identified.}
	\Description{}
	\label{fig-graph}
\end{figure}

For each toolpath curve $c$ in $H_{i}$ split into multiple edges, we can identify if an edge $e = (S,E)$ is supported by looking at the palisade edge $e_{L}$ incident to $E$ that is to the left of $e$. $e$ is supported if and only if the intrinsic direction of $e_{L}$ is leaving $E$ (Figure~\ref{fig-graph}a). For $e$, the \emph{next valid edge} is the edge incident to $E$ with a referenced direction leaving $E$ (Figure~\ref{fig-graph}b). Add $e$ and its next valid edge to $H_{i}^{\text{collected}}$.

Later all collected edges can be joined to form closed, non-intersecting curves. To prove so, it suffices to show that every vertex in $G$ has exactly two incident edges with consistent directions collected. For a start vertex $S$ (Figure~\ref{fig-graph}c), depending on the referenced direction of the palisade at $S$, we can find the neighboring supported edge pointing at that palisade and its next valid edge will be pointing at $S$. Since each vertex has one incident toolpath edge collected, and the number of toolpath edges and palisade edges collected are identical, no other edges other than those highlighted will be collected. The edges can thus be joined at every vertex to form closed curves only. Naturally they are intersection-free.

In our implementation, the selection of these edges and the joining is done in a single pass of depth-first search (DFS) on $G$.

\subsubsection{Represent fully unsupported curves for consistent orientation}\label{sec-replace-unsupported-curves}

\begin{figure}
	\centering
	\includegraphics[width=85mm]{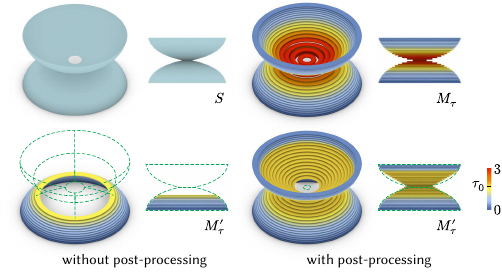}
	\caption{A fully unsupported curve may cause the upper body in subsequent layers to disappear after optimization. With the proposed post-processing, the curve can be represented by a palisade curve to preserve the upper body.}
	\Description{}
	\label{fig-vanish-body}
\end{figure}

Intuitively, if a curve is fully unsupported, it should be abandoned. However, we prefer to ``represent'' them somehow using palisade curves in the optimization results because losing a complete curve may (1) cause the upper body in subsequent layers to disappear completely, which is not desired (example in Figure~\ref{fig-vanish-body}); and (2) break the consistent orientations that form regions, causing problems in counting the winding number.

\begin{figure}
	\centering
	\includegraphics[width=85mm]{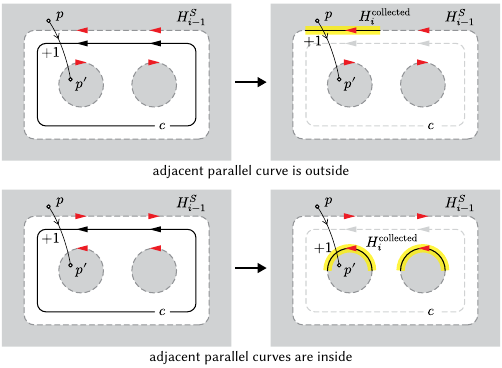}
	\caption{A fully unsupported curve can be represented using adjacent palisade curves with parallel referenced directions.}
	\Description{}
	\label{fig-pending-curve-representation}
\end{figure}

For a curve $c$ in $H_{i}^{\text{pending}}$, we find its adjacent curve(s) in $\partial H_{i - 1}^{S}$ (Figure~\ref{fig-pending-curve-representation}).
They may include one curve outside $c$ and some curves inside $c$. Since $c$ is out of the region of $H_{i - 1}^{S}$, the two sides have opposite referenced directions. (Here we assume $c$ is the only curve in $H_{i}^{\text{pending}}$ between them. If not, the consistent orientation has to be restored in the cleaning of the closed curve tree in Section~\ref{sec-clean-final-result}.) Consider a trajectory $pp'$ that passes through the three layers. To maintain the change in the winding number in $H_{i}^{\text{collected}}$ after the removal of $c$, we find the palisade curves of the adjacent side with parallel referenced directions. For each palisade, if at least one edge from it is already in $H_{i}^{\text{collected}}$, we assume the winding number is preserved by this. Otherwise, we collect the entire palisade in its referenced direction to represent the unsupported curve $c$. If no parallel palisade curve exists, we have to simply abandon $c$. This post-processing is in accordance with Eq.~(\ref{eq-local-modification-2}).


\Needspace{12\baselineskip}
\begin{wrapfigure}{r}{27.5mm}
	\centering
	\includegraphics[width=27.5mm]{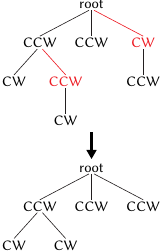}
\end{wrapfigure}
\subsubsection{Clean final result for consistent orientation}\label{sec-clean-final-result}
\begin{figure}
	\centering
	\includegraphics[width=85mm]{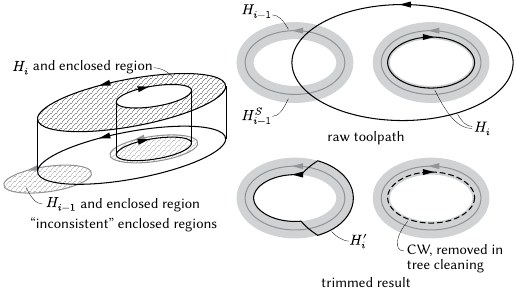}
	\caption{Closed curve tree cleaning will remove curves only when regions are ``inconsistent'' across adjacent layers.}
	\Description{}
	\label{fig-tree-cleaning-needed}
\end{figure}

While the consistent orientation should be preserved in the previous steps, it may be broken for corner cases where the two adjacent toolpath layers have little correlation. To ensure consistent orientation, we build the hierarchical structure of enclosure formed by the closed curves in $H_{i}^{\text{collected}}$, known as the \emph{closed curve tree}. The curves, from exterior to interior, should be alternately oriented CCW and CW. To clean the tree, we traverse it and skip those breaking the rule. The remaining curves, with consistent orientation, are collected as the final result $H'_{i}$. In Section~\ref{sec-results-and-discussion}, none of the collected examples had their curves removed in this cleaning. The cleaning only changes the result for extreme corner cases, one of which is illustrated in Figure~\ref{fig-tree-cleaning-needed}, where a fully supported curve changes from a CW inner wall to a CCW outer wall and has to be removed.

\begin{figure*}
	\centering
	\includegraphics[width=179mm]{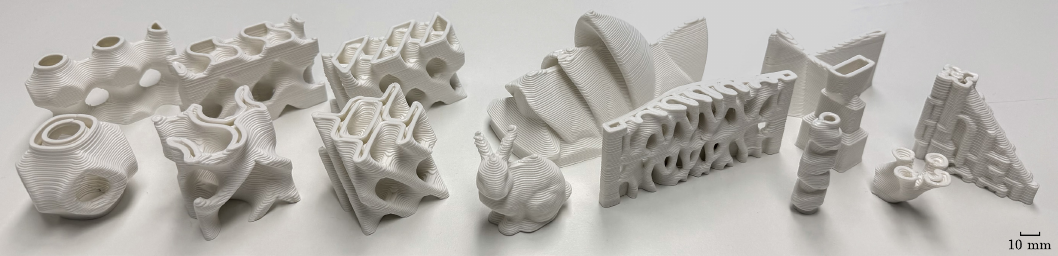}
	\caption{Optimized shell printing toolpaths can be printed without overhang issues. Examples printed using PLA with a $0.7\times 2.1 \textrm{ mm}$ extrusion profile.}
	\Description{}
	\label{fig-PLA-photo}
\end{figure*}


\definecolor{TypstGreen}{HTML}{2ecc40}
\definecolor{TypstRed}{HTML}{FF4136}

\ExplSyntaxOn
\cs_new:Npn \tbl_green_from_lighten_pct:n #1
{
	\fp_eval:n { round(100 - min(100, max(0, #1)), 2) }
}
\cs_new:Npn \tbl_pct_chg:n #1 { \tbl_green_from_lighten_pct:n { abs(#1) * 40 } }
\cs_new:Npn \tbl_pct_iou:n #1 { \tbl_green_from_lighten_pct:n { (1 - #1) * 400 } }
\cs_new:Npn \tbl_pct_time:n #1 { \tbl_green_from_lighten_pct:n { #1 } }
\cs_new:Npn \tbl_pct_tpk:n #1 { \tbl_green_from_lighten_pct:n { #1 * 15 } }
\NewDocumentCommand{\heatChg}{m m}{\cellcolor{TypstGreen!\tbl_pct_chg:n{#1}} #2}
\NewDocumentCommand{\heatIou}{m}{\cellcolor{TypstGreen!\tbl_pct_iou:n{#1}} #1}
\NewDocumentCommand{\heatTime}{m m}{\cellcolor{TypstGreen!\tbl_pct_time:n{#1}} #2}
\NewDocumentCommand{\heatTpk}{m m}{\cellcolor{TypstGreen!\tbl_pct_tpk:n{#1}} #2}
\ExplSyntaxOff

\providecommand{\Yes}{\ding{51}}
\providecommand{\No}{\ding{55}}
\newcommand{\YesRed}{\textcolor{TypstRed}{\ding{51}}}
\newcommand{\NoGreen}{\textcolor{TypstGreen}{\ding{55}}}

\begin{table*}[thbp]
	\centering
	\caption{Optimization results show high similarity scores in the perceived area.}
	\label{tab-examples}
	\small
	\setlength{\tabcolsep}{4pt}
	\renewcommand{\arraystretch}{1.05}
	\resizebox{0.8\textwidth}{!}{%
		\begin{tabular}{c c l c c c c c c c c c c c c}
			\toprule
			Fig.                               &      & Example               & \#L & \#C & \#P   & $\tau_0$ & US       & CL       & Length change                & Length IoU      & Area change                  & Area IoU        & Time                    & Time / 1k pts               \\
			\midrule
			\ref{fig-examples}                 & (1)  & Pipe                  & 24  & 24  & 4835  & 1.5      & \NoGreen & \NoGreen & \heatChg{-3.95}{-3.95\%}     & \heatIou{0.801} & \heatChg{-7.26}{-7.26\%}     & \heatIou{0.920} & \heatTime{14}{14 ms}    & \heatTpk{2.896}{2.896 ms}   \\
			                                   &      &                       &     &     &       & 1.1      & \NoGreen & \NoGreen & \heatChg{-7.75}{-7.75\%}     & \heatIou{0.667} & \heatChg{-14.92}{-14.92\%}   & \heatIou{0.830} & \heatTime{11}{11 ms}    & \heatTpk{2.275}{2.275 ms}   \\
			                                   &      &                       &     &     &       & 0.7      & \NoGreen & \NoGreen & \heatChg{-17.05}{-17.05\%}   & \heatIou{0.483} & \heatChg{-28.31}{-28.31\%}   & \heatIou{0.674} & \heatTime{9}{9 ms}      & \heatTpk{1.861}{1.861 ms}   \\

			                                   & (2)  & Bunny                 & 64  & 82  & 6973  & 1.5      & \NoGreen & \NoGreen & \heatChg{-1.70}{-1.70\%}     & \heatIou{0.820} & \heatChg{-0.50}{-0.50\%}     & \heatIou{0.982} & \heatTime{32}{32 ms}    & \heatTpk{4.589}{4.589 ms}   \\
			                                   &      &                       &     &     &       & 1.0      & \NoGreen & \NoGreen & \heatChg{-3.09}{-3.09\%}     & \heatIou{0.605} & \heatChg{-1.31}{-1.31\%}     & \heatIou{0.939} & \heatTime{30}{30 ms}    & \heatTpk{4.302}{4.302 ms}   \\

			\ref{fig-examples-2}               & (3)  & Skewer 1              & 74  & 74  & 370   & 1.0      & \YesRed  & \NoGreen & \heatChg{-0.10}{-0.10\%}     & \heatIou{0.779} & \heatChg{0.31}{+0.31\%}      & \heatIou{0.940} & \heatTime{6}{6 ms}      & \heatTpk{16.216}{16.216 ms} \\
			                                   & (4)  & Skewer 2              & 73  & 77  & 5599  & 1.0      & \NoGreen & \NoGreen & \heatChg{-3.29}{-3.29\%}     & \heatIou{0.583} & \heatChg{-2.03}{-2.03\%}     & \heatIou{0.925} & \heatTime{21}{21 ms}    & \heatTpk{3.751}{3.751 ms}   \\
			                                   & (5)  & Sculpture             & 68  & 138 & 76439 & 1.0      & \NoGreen & \NoGreen & \heatChg{-0.09}{-0.09\%}     & \heatIou{0.831} & \heatChg{1.54}{+1.54\%}      & \heatIou{0.971} & \heatTime{174}{174 ms}  & \heatTpk{2.276}{2.276 ms}   \\
			                                   & (6)  & Sydney Opera House    & 91  & 121 & 15958 & 1.0      & \YesRed  & \NoGreen & \heatChg{3.18}{+3.18\%}      & \heatIou{0.660} & \heatChg{2.58}{+2.58\%}      & \heatIou{0.967} & \heatTime{73}{73 ms}    & \heatTpk{4.575}{4.575 ms}   \\
			                                   & (7)  & St. Mary's Cathedral  & 68  & 86  & 10941 & 0.5      & \NoGreen & \NoGreen & \heatChg{2.92}{+2.92\%}      & \heatIou{0.734} & \heatChg{3.96}{+3.96\%}      & \heatIou{0.962} & \heatTime{42}{42 ms}    & \heatTpk{3.839}{3.839 ms}   \\

			\ref{fig-examples-tpms}            & (8)  & Schwarz P TPMS Unit   & 60  & 164 & 14000 & 1.0      & \NoGreen & \NoGreen & \heatChg{1.54}{+1.54\%}      & \heatIou{0.736} & \heatChg{-0.87}{-0.87\%}     & \heatIou{0.948} & \heatTime{74}{74 ms}    & \heatTpk{5.286}{5.286 ms}   \\
			                                   & (9)  & Gyroid TPMS Unit      & 60  & 148 & 20866 & 1.0      & \NoGreen & \NoGreen & \heatChg{0.57}{+0.57\%}      & \heatIou{0.641} & \heatChg{-3.01}{-3.01\%}     & \heatIou{0.866} & \heatTime{94}{94 ms}    & \heatTpk{4.505}{4.505 ms}   \\
			                                   & (10) & Diamond TPMS Unit     & 60  & 208 & 29144 & 1.0      & \NoGreen & \NoGreen & \heatChg{5.97}{+5.97\%}      & \heatIou{0.672} & \heatChg{2.18}{+2.18\%}      & \heatIou{0.890} & \heatTime{152}{152 ms}  & \heatTpk{5.215}{5.215 ms}   \\

			                                   & (11) & Schwarz P TPMS Infill & 62  & 124 & 11200 & 1.0      & \NoGreen & \NoGreen & \heatChg{2.05}{+2.05\%}      & \heatIou{0.799} & \heatChg{0.33}{+0.33\%}      & \heatIou{0.987} & \heatTime{47}{47 ms}    & \heatTpk{4.196}{4.196 ms}   \\
			                                   & (12) & Gyroid TPMS Infill    & 62  & 176 & 19733 & 1.0      & \NoGreen & \NoGreen & \heatChg{1.81}{+1.81\%}      & \heatIou{0.766} & \heatChg{-0.60}{-0.60\%}     & \heatIou{0.970} & \heatTime{94}{94 ms}    & \heatTpk{4.764}{4.764 ms}   \\
			                                   & (13) & Diamond TPMS Infill   & 62  & 236 & 26362 & 1.0      & \NoGreen & \NoGreen & \heatChg{5.79}{+5.79\%}      & \heatIou{0.743} & \heatChg{-0.52}{-0.52\%}     & \heatIou{0.963} & \heatTime{154}{154 ms}  & \heatTpk{5.842}{5.842 ms}   \\

			\ref{fig-examples-topoopt}         & (14) & TopoOpt Design        & 36  & 77  & 17987 & 1.0      & \NoGreen & \NoGreen & \heatChg{-0.62}{-0.62\%}     & \heatIou{0.767} & \heatChg{3.23}{+3.23\%}      & \heatIou{0.947} & \heatTime{76}{76 ms}    & \heatTpk{4.225}{4.225 ms}   \\
			\ref{fig-examples-artificial-reef} & (15) & Artificial Reef       & 66  & 230 & 52444 & 1.0      & \NoGreen & \NoGreen & \heatChg{0.04}{+0.04\%}      & \heatIou{0.876} & \heatChg{4.09}{+4.09\%}      & \heatIou{0.945} & \heatTime{419}{419 ms}  & \heatTpk{7.989}{7.989 ms}   \\

			                                   &      & (Average)             &     &     &       &          &          &          & \heatChg{3.42}{$\pm 3.42\%$} & \heatIou{0.720} & \heatChg{4.31}{$\pm 4.31\%$} & \heatIou{0.924} & \heatTime{84.56}{85 ms} & \heatTpk{4.922}{4.922 ms}   \\
			\bottomrule
		\end{tabular}%
	}

	\vspace{0.6ex}
	\parbox{0.8\textwidth}{\footnotesize\raggedright
		\#L: number of layers.
		\#C: number of curves.
		\#P: number of points.
		US: unsupported curves in $H_i^{\mathrm{pending}}$ are represented by additional palisades of $\partial H_{i-1}^{S}$.
		CL: some curves are removed in the final cleaning.
		Time is recorded on a laptop with an Intel Core i9-13950HX CPU and 32\,GB of RAM.}
\end{table*}

\begin{figure}
	\centering
	\includegraphics[width=85mm]{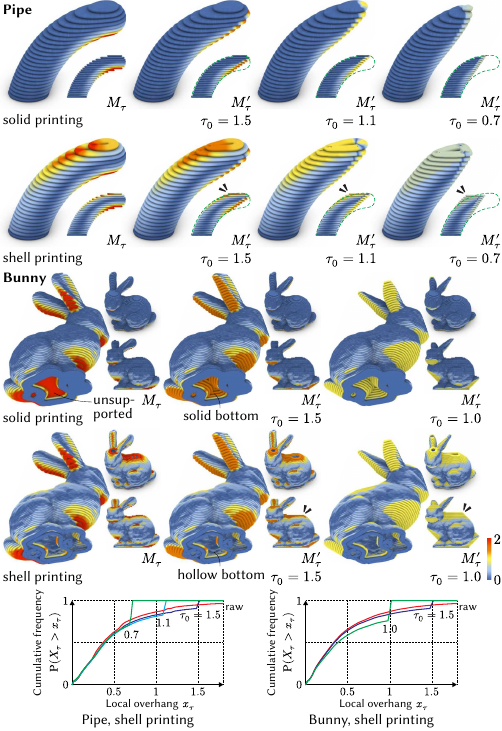}
	\caption{Optimizations on solid and shell printing of the same geometry render different results.}
	\Description{}
	\label{fig-examples}
\end{figure}

\subsection{Implementation}

The proposed algorithm has been implemented in C\# as
a plugin
for Grasshopper® in Rhino®. Polygon offsetting is done with Clipper2 \cite{clipper2}. (All curves are polygons in our implementation. The algorithm applies to any type of curve if the respective offsetting method is available.) Curve intersections, curve-point inclusion, and orientations are solved with native RhinoCommon methods \cite{RhinoCom27:online}. The bottleneck of the computation time is the polygon offsetting and curve-related methods. The rest of the algorithm is simple graph search and uses negligible time.

\section{Results and applications}\label{sec-results-and-discussion}

We present a series of results in this section and discuss the effects and applications of the algorithm. After optimization, short curves ($<5\times$ layer thickness) are ruled unprintable and culled. Physical experiments verify the printability of the optimized shell printing toolpaths in PLA (Figures~\ref{fig-teaser} and \ref{fig-PLA-photo}), clay (Figure~\ref{fig-bunny-clay}), and concrete (Figure~\ref{fig-examples-artificial-reef}).

\subsection{Shell printing parameters}

Table~\ref{tab-examples} summarizes the shell printing examples and their parameters. To quantify the complexity of the inputs, we collect the number of layers, curves, and polyline points of each example. To assess the change of the optimization, we document the length/area change and the intersection over union (IoU) of both the toolpath curves and the regions enclosed.

The change in length reflects the amount of material added/saved and the change in the area suggests the change in the perceived volume. The IoU is a similarity metric defined as the ratio of the intersection curve/area to the union curve/area of the input and output sets. The numerator of the length IoU is the total length of the segments in $H_{i}$ that are supported and preserved. While the average length IoU for the examples is $0.72$, the average area IoU is $0.92$, meaning the optimization induces a conservative change to the region areas, thereby preserving the original shape.

The running time of the optimization algorithm is below 200 ms except for the largest example on the test setup and scales with the size of the input (primarily the number of points, on average $5$ms per $1000$ points).

\subsection{Solid vs. shell printing}

In the first two sets of examples (Figure~\ref{fig-examples}) we compare the optimization for solid and shell printing. In shell printing, the connection scheme may create volume that protrudes out of the input geometry. Therefore, although the method initially adopts a ``subtractive'' material strategy, we are also ``adding'' material when needed. Unlike in solid printing, we do not necessarily have $V' \subseteq V$ in shell printing.

\begin{figure}
	\centering
	\includegraphics[width=85mm]{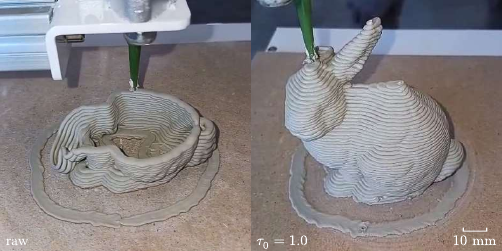}
	\caption{Shell printing of the bunny in clay is made possible by the optimization. Examples printed using EAZAO Potter with a 1 × 3 mm extrusion profile.}
	\Description{}
	\label{fig-bunny-clay}
\end{figure}

\begin{figure}
	\centering
	\includegraphics[width=85mm]{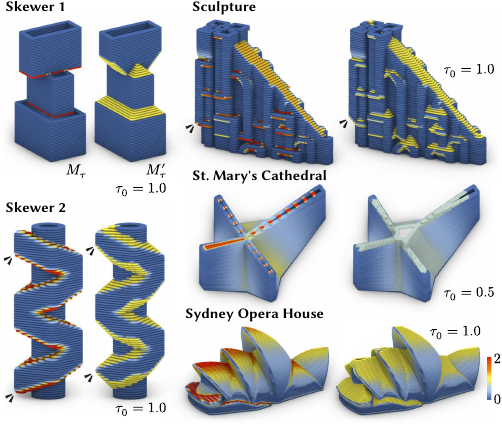}
	\caption{Shell printing optimization creates blending and smoothing effects.
	}
	\Description{}
	\label{fig-examples-2}
\end{figure}

\begin{figure*}
	\centering
	\includegraphics[width=179mm]{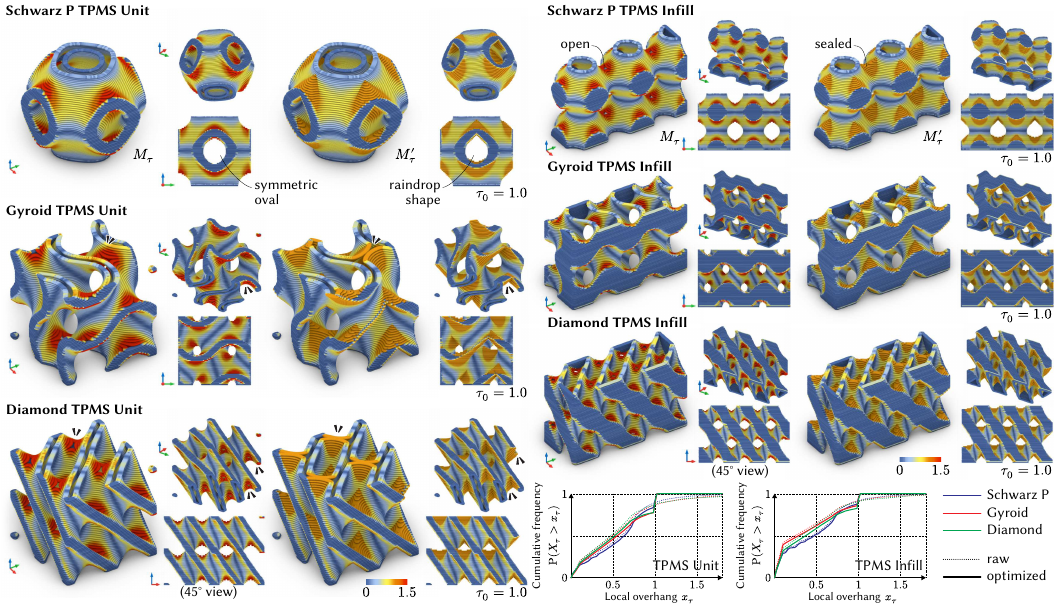}
	\caption{Triply periodic minimal surface (TPMS) toolpaths are optimized to be buildable and sealed.}
	\Description{}
	\label{fig-examples-tpms}
\end{figure*}

\begin{figure*}
	\centering
	\includegraphics[width=179mm]{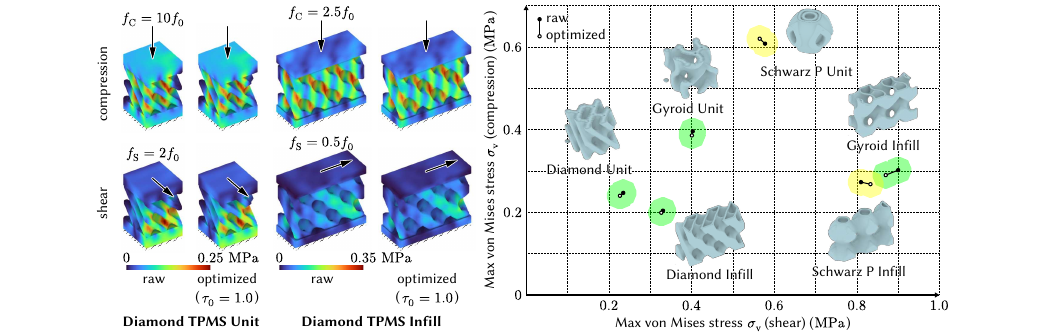}
	\caption{FEA shows that TPMS prototypes optimized for buildability maintain structural performance. (To incorporate self weight, $f_{0}$ is defined as the weight of concrete with the volume of the bounding box of the TPMS samples, and the compression and shear forces are applied proportionally.)}
	\Description{}
	\label{fig-examples-tpms-fe}
\end{figure*}

Another difference is the effect on inner walls that represent openings of the surface. For Bunny, in solid printing, the bottom opening creates a large unsupported area and, by removing some materials, the optimized version uses a staircase-style closing to support the solid bottom. In shell printing, the hollow bottom is deemed buildable and thus unchanged.

Figure~\ref{fig-examples} also shows the robust effect of different overhang limits $\tau_{0}$. The cumulative frequency graphs show that the result is bounded by the overhang limit $\tau_{0}$. Depending on the shape, the distribution of lower overhangs would also change. Physical experiment with clay (Figure~\ref{fig-bunny-clay}) confirms the buildability of the optimized shell printing toolpath.

The rest of this section focuses on shell printing.

\subsection{Blending and smoothing effects}\label{sec-blending-and-smoothing-effects}

Figure~\ref{fig-examples-2} provides more shell printing examples from which we observe \emph{blending} and \emph{smoothing} effects. Skewer~1 gives a minimal example of a completely unbuildable design. The result creates uniform blended connection between the discontinuous blocks. Skewer~2 and Sculpture show how sharp corners are smoothed out and cantilever features are trimmed (marked).
Sydney Opera House offers a smooth shell surface example where the optimization ensures buildability while preserving curvature continuity.
St. Mary's Cathedral shows that a creased hyperbolic design sees transitional fillets at its creases to remove extreme inward overhang.

\subsection{Triply periodic minimal surface (TPMS) units and infills}

Triply periodic minimal surface (TPMS) structures are known for their structural efficiency, maximized area, and lightweight properties \cite{Yu-2025-3DP-Lattice}. While such geometries might be desirable in high-performance large-format 3D printed structures, their extreme overhangs have required external support schemes or custom post-processing and thus limited their application. Here we test the proposed method on TPMS units and infills to demonstrate its potential for expanding the application of such geometries in large-format 3D printing.

Let $\ell$ denote the wavelength of the TPMS. Figure~\ref{fig-examples-tpms} shows $\ell \times \ell \times \ell$ TPMS units of three types: Schwarz P, Diamond, and Gyroid. To ensure curve closure, the units are double-sided. In addition, single-sided TPMS patterns that act as infills or supports are included as in an $\ell \times 2.5\ell \times 1.5\ell$ box. We adopted a $\tau_{0} = 1.0$ ($\theta_{0} = 45{^\circ}$) uniform overhang limit. In both units and infills the overhangs are successfully optimized to be under the limit. The method replaced the red extreme overhang zones with long, continuous ramps without changing the porous topology. It also sealed the open holes caused by slicing. The original symmetric oval holes are transformed to have raindrop shapes, the classic void shape in porous buildable designs \citep[see][]{Guo-2017-3DP-Topological-Optimization}.

Finite element analysis (FEA) using Intact.Simulation \cite{Intact-2026-Simulation} was performed on the TPMS units and infills to assess the structural performance of the optimized results (Figure~\ref{fig-examples-tpms-fe}). While the optimized results deviate from the exact minimal surfaces, the FEA result shows that Gyroid and Diamond prototypes see reduction in the max von Mises stress recorded for both compression and shear, suggesting an improvement in structural performance. This might be due to the fact that the raw sliced toolpath has thin or open walls at the horizontal zones, while the optimized results are more continuous and uniform in material thickness. Schwarz P prototypes see mixed changes in compression and shear stress.

\subsection{Compatibility with topology optimization}

\begin{figure}
	\centering
	\includegraphics[width=85mm]{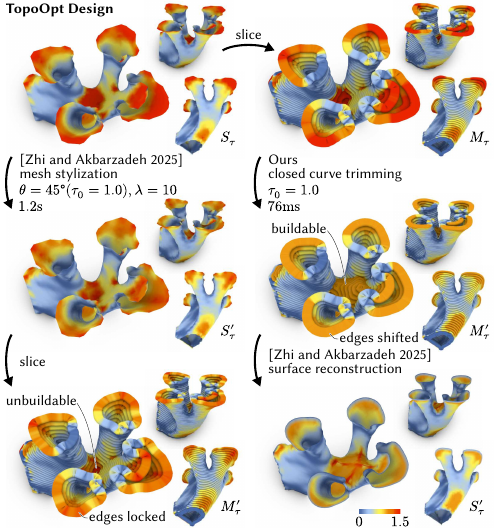}
	\caption{The optimization method is compatible with structures designed by topology optimization. Compared with the iterative mesh stylization method \cite{Zhi-2025-Overhang-2}, the proposed method is non-iterative and does not leave unbuildable concave zones.}
	\Description{}
	\label{fig-examples-topoopt}
\end{figure}

To show the compatibility of the optimization algorithm with existing design/form-finding methods and engage with the precedent studies of overhang-optimized design/form-finding methods (Table~\ref{tab-current-methods}), we demonstrate the example of a branching structure designed by topology optimization (Figure~\ref{fig-examples-topoopt}). In \citet{Zhi-2025-Overhang-2}, this geometry was optimized as a mesh using normal-driven stylization. The upper bound of the overhang was not guaranteed, and the result involves an unbuildable concave zone in the middle. The proposed method optimized the sliced toolpath of the original mesh. The optimized overhang is bounded by the overhang limit, and the concave zone is replaced by a continuous, buildable ramp. Using surface reconstruction \cite{Zhi-2025-Overhang-2}, the optimized toolpath can be converted back to a mesh for further processing and fabrication.

\subsection{Internal overhang reduction of a complex toolpath}

\begin{figure}
	\centering
	\includegraphics[width=85mm]{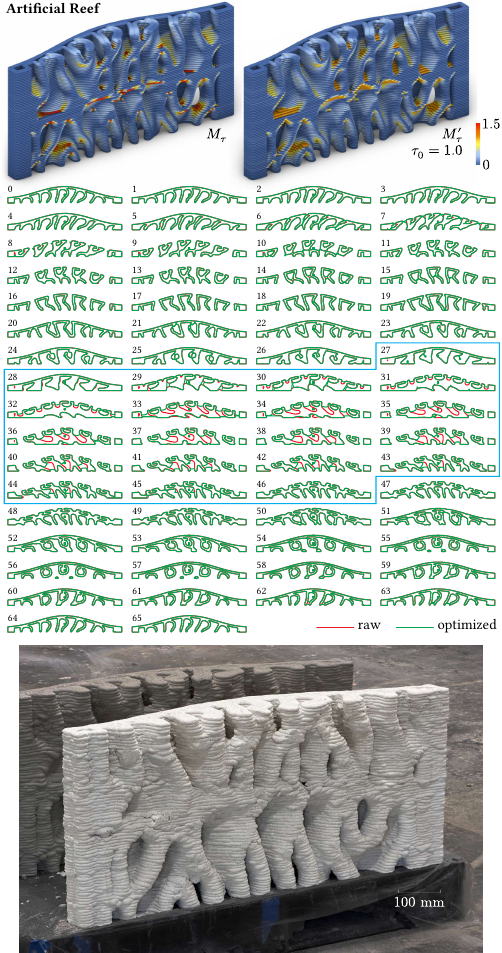}
	\caption{The optimization of a porous artificial reef structure shows that the algorithm only modifies extreme overhang regions when necessary and converges back to the original toolpath afterwards. The example is printed using 3D Potter and Sika concrete with a $7\times 21 \textrm{ mm}$ extrusion profile.}
	\Description{}
	\label{fig-examples-artificial-reef}
\end{figure}

The last example illustrates how the method optimizes a design with a complex internal structure while preserving its external shape. The porous structure (Figure~\ref{fig-examples-artificial-reef}) is designed by embedding a coral-reef-like pattern with branching and folding features in a bounding geometry of a ramp. The overall buildability is fine since the infill pattern is guided to be parallel with the $+ z$ direction, but extreme overhang is observed in bridges, cantilever parts, and steep ramps. The optimized toolpath remains almost unchanged in layers 0--26, and sees critical change in layers 27--46 where the extreme overhang regions are replaced by buildable ramps (red vs. green). Finally, in layers 47--65, the optimized layers converge back to the original state. The boundary shape is preserved for the block to act as a tessellation tile with a continuous pattern.

\begin{figure}
	\centering
	\includegraphics[width=85mm]{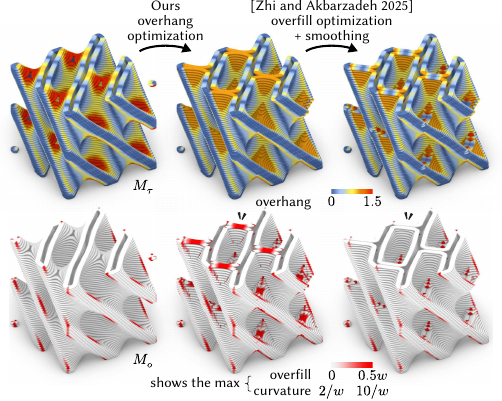}
	\caption{The overfill optimization and smoothing method by \citet{Zhi-2025-Overhang-2} reduces the overfilled areas in the overhang-optimized toolpath while introducing little area that exceeds the overhang limit.}
	\Description{}
	\label{fig-overfill-corrected-example}
\end{figure}

\section{Limitations and potential solutions}\label{sec-limitations-and-solutions}

\subsection{Overfilled parts}

Since the overhang optimization method adopts a purely graph-based view, it might create overfilled areas and sharp corners that compromise the print quality. We suggest performing overfill optimization and smoothing to post-process the toolpath to enhance print quality (Figure~\ref{fig-overfill-corrected-example}). In Appendix~\ref{apd-overfill-optimization}, a sketchy analysis shows that the overhang added in this process is small.


\subsection{Planar openings on the print}

For functional and aesthetic reasons, sometimes we prefer the print result to be watertight without planar openings. The general idea is that every little segment of curve $s$ in $H_{i-1}$ should be either (1) deemed as closed or (2) ``inherited'' by another segment in $H_{i}$ that is supported by $s$. If not, certain palisade curves should be added to $H_{i}$. To strictly enforce this without breaking the manifold topology would require point-level offset and dependency tracking and is not compatible with the design of the proposed algorithm.

\Needspace{8\baselineskip}
\begin{wrapfigure}{r}{28mm}
	\centering
	\includegraphics[width=28mm]{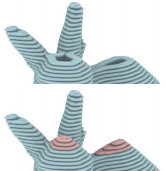}
	\Description{}
\end{wrapfigure}

\Needspace{8\baselineskip}
Here we provide a band-aid solution that only works for certain simple geometries without holes in their sliced regions. Before the final cleaning, we collect any palisade curve $c$ that is the result of offsetting some curve in $H_{i-1}$ to its left if none of $c$'s segments are already collected.
If $c$ represents a planar opening it will be preserved in the final cleaning. If $c$ is redundant, in the closed curve tree it will be wrapped by another curve that is more generic and thus removed. This hack would effectively shrink CCW planar openings in the subsequent layers until they are sealed and therefore no curve is further created when offsetting to the left. Figure~\ref{fig-teaser} shows the print result with this method applied.

\subsection{Open curves}

To expand the algorithm to accommodate open curves that represent a manifold but open surface, there are two potential directions. The first is to close the open curves by adding temporary segments, run overhang optimization, and open up the results. However, to solve such a scheme without creating intersections is non-trivial and different schemes might give different results. Moreover, the temporary segments might be significantly modified and relied upon, thereby making it impossible to recover the open curves in a buildable way.

Another direction is to ``thicken'' all curves to be double-sided, run overhang optimization, and solve the linear medial axis of the results. However, the regions enclosed by the double-sided walls might deform to involve branching shapes, making the linear medial axis complex to solve. Furthermore, the medial axis itself does not have a given curve direction and is not guaranteed to be buildable.

It is thus an open question whether the proposed algorithm can be expanded to apply to open curves.

\section{Conclusion}
We have presented the closed curve trimming method of toolpath overhang optimization for support-free planar shell printing. The method applies to arbitrary manifold, closed shapes and guarantees by construction that the resulting toolpath curves stay within the overhang limit while preserving their orientation and preventing intersections. Shown in the case studies, the shapes are largely preserved and the shell characteristics are maintained.

Its main limitation is that it only applies to closed curves, and that the results are limited by the bottom-up process. If a similar decision protocol can be designed for the top-down direction to add/modify the bottom layers, the algorithm can potentially achieve more balanced results.



\begin{acks}
	This work was supported by the National Science Foundation Future Eco Manufacturing Research Grant (NSF, FMRGCMMI2037097). 

	We acknowledge the designers/authors of the 3D models used in this paper, including
	Bunny \cite{bunny},
	Skewer~2 based on \citet{Mitropoulou-2020-Non-Parallel-3DP},
	Sydney Opera House designed by Jørn Utzon, model by \citet{OperaHou29:online},
	St. Mary's Cathedral designed by Kenzo Tange,
	TopoOpt Design \cite{Zhi-2025-Overhang-2},
	and
	Artificial Reef \cite{Machado-Romero-2026-Artificial-Reef}. We thank David Chang and Eazao for the contribution in clay printing experiments. We also thank Amir Motavaselian, Fima Ivanov, and Hadi Rashidi for participating in the concrete printing experiments.
\end{acks}


\bibliographystyle{ACM-Reference-Format}
\bibliography{yefan_latex}

\appendix

\section{Touching points in 2D volume boundaries}\label{apd-touching-boundary}

\Needspace{9\baselineskip}
\begin{wrapfigure}{r}{39mm}
	\centering
	\includegraphics[width=39mm]{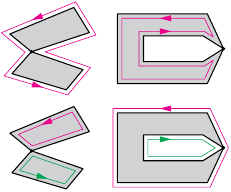}
\end{wrapfigure}
A touching point in the 2D volume is a point whose neighborhood is divided into multiple parts by the boundaries. The point is either on a self-intersecting boundary curve or on multiple boundary curves that intersect at that point. To form $G$, we intersect every pair of curves from $\partial H_{i - 1}^{S}$ and $H_{i}$, and record the intersections on both curves using parameters. If a touching point happens to be an intersection, it brings ambiguity in sorting the intersections along a curve and will break the property required in selecting the supported segments and their next valid edges (Figure~\ref{fig-graph}).

In extreme cases, a touching point can be incident to infinitely many edges, making it impractical to minimally adjust the boundary curves before solving $G$ to remove the touching point without creating new intersections. Therefore, in our implementation, we simply assume there are no touching points that happen to be an intersection between $\partial H_{i - 1}^{S}$ and $H_{i}$.

\Needspace{14\baselineskip}
\section{Overlapping singularities in $G$}\label{apd-overlap}

\Needspace{12\baselineskip}
\begin{wrapfigure}{r}{52mm}
	\centering
	\includegraphics[width=52mm]{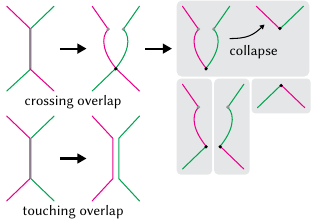}
	\Description{}
\end{wrapfigure}
When solving $G$, an overlapping segment instead of a point intersection may occur, especially when the original ramp already has an overhang of $\tau_{0}$.
For a crossing overlap, we disentangle one end and place two pseudo vertices without creating a new intersection. The pseudo vertices mark the overlapping edges which, if collected consecutively, will be collapsed.

When the overlap is touching, we simply ignore it.

\section{Overhang added in overfill optimization}\label{apd-overfill-optimization}

Overfill happens when the distance between two segments is less than the extrusion width $w$.
Assuming the original toolpath has no overfill, we first optimize its overhang, then perform overfill optimization to reduce the overfilled area.

\Needspace{7\baselineskip}
\begin{wrapfigure}{r}{43mm}
	\centering
	\includegraphics[width=43mm]{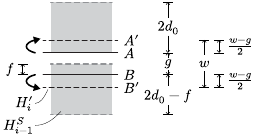}
	\Description{}
\end{wrapfigure}
Here we estimate the overhang added in overfill optimization that exceeds the buildable limit.
Consider two parallel segments $A$ and $B$ in $H'_i$ that are too close to each other after overhang optimization, with a distance $g<w$. It must be the result of one or both segments being offset \emph{inward} from the raw state. Say $A$ is the result of offsetting and is now on $\partial H_{i-1}^S$. Let $f$ denote the distance between $B$ and $\partial H_{i-1}^S$. We have $0\leq f <g$.
Overfill optimization then pulls $A$ and $B$ apart symmetrically to $A'$ and $B'$ to have a clearance of $w$.

For now, assume $H_{i-1}$ stays unchanged.
If $B'$ is outside $\partial H_{i-1}^S$, the added overhang will be
\begin{align}
	e_1=\frac{w-g}{2}-(2d_0-f)< \frac{w+g}{2}-2 d_0< w- 2d_0.
\end{align}
If $d_0\geq w/2$, no overhang is added and the optimized layer is still buildable.

Now consider when $H_{i-1}$ is also altered by overfill optimization. The part that supports $B'$ will most likely be pulled to the same direction, making $e_1$ smaller.

Overfill at sharp corners is resolved by smoothing the corners until they reach the target corner radius, which is typically $r = w/2$. After smoothing, the span of the arc is bounded by
\begin{align}
	e_2 \leq 2r= w.
\end{align}
The largest added overhang will be $e_2/2$ at the mid-span.
Since the arc bridge is supported on both ends, it can be considered buildable in most printing scenarios.





\end{document}